\documentclass[aps,prl,showpacs,twocolumn,floats,superscriptaddress]{revtex4}
\usepackage{graphicx}
\usepackage{dcolumn}
\usepackage{bm}
\usepackage{color}
\usepackage{amssymb}

\begin{document}
%%%%%%%%%%%%%%%%%%%%%%%%%%%%%%%%%%%%%%
%%%%%%%%%%%%%%%%%%%%%%%%%
\author{\c{S}.~G. S\"{o}yler}
\affiliation{The Abdus Salam International Centre for Theoretical Physics,
Strada Costiera 11, I-34151 Trieste, Italy}

\author{M. Kiselev}
\affiliation{The Abdus Salam International Centre for Theoretical Physics,
Strada Costiera 11, I-34151 Trieste, Italy}

\author{N. V. Prokofev}
\affiliation{Department of Physics,
University of Massachusetts, Amherst, MA
01003, USA}
\affiliation{Russian Research Center
``Kurchatov Institute'', 123182 Moscow,
Russia}

\author{B. V. Svistunov}
\affiliation{Department of Physics,
University of Massachusetts, Amherst, MA
01003, USA}
\affiliation{Russian Research Center
``Kurchatov Institute'', 123182 Moscow,
Russia}
%%%%%%%%%%%%%%%%%%%%%%%%%
\title{Phase Diagram of Commensurate Two-Dimensional Disordered Bose Hubbard Model}
%%%%%%%%%%%%%%%%%%%%%%%%%%%%%%%%%%%%%%

\begin{abstract}
We establish the full groundstate phase diagram of disordered Bose-Hubbard model in
two-dimensions at unity filling factor via quantum Monte Carlo simulations.
Similarly to the three-dimensional case we observe extended superfluid regions
persisting up to extremely large values of disorder and interaction strength which,
however, have small superfluid fractions and thus low transition temperatures.
In the vicinity of the superfluid--insulator transition of the pure system,
we observe an unexpectedly weak---almost not resolvable---sensitivity of the critical interaction to the strength of (weak) disorder.
\end{abstract}


\maketitle

%67.85.Hj	Bose-Einstein condensates in optical potentials
% 67.85.-d	Ultracold gases, trapped gases
% 64.70.Tg	Quantum phase transitions

%%%%%%%%%%%%%%%%%%%%%%%%%%%%%%%%%%%%%%%%
%%%%%%%%%%%%%%%%%%%

Disordered systems keep attracting a lot of attention as they reveal rich
and nontrivial physics of interplay between interaction effects and localization
\cite{Giamarchi,Fisher}. Bosons are particularly interesting because
non-interacting particles are always localized in the region of space which
corresponds to the global minimum of the disorder potential, i.e.
the limit of weak disorder and weak interactions is already singular.
The study of the so called `dirty boson problem' was first
prompted by the disappearance of superfluidity of $^4$He in porous
media (vycor) \cite{He4_vycor_1_2_3}.
Other condensed matter systems of interest include thin superconducting films \cite{sc_film}, Josephson junction arrays \cite{josephson_junction},
superfluid helium films on substrates \cite{He_thinfilms_1_2}, etc.

More recently, the realization of ultracold atoms in optical lattices
\cite{Jaksch, Greiner} paved the road to studies of strongly correlated systems
in a controlled way, and to using them as emulators for
condensed matter models. In the presence of disorder, most cold atomic systems are
described by the Bose-Hubbard model with disorder in the on-site potential.

Control over disorder has been implemented in optical lattices in several
ways such as using bicromatic lattices, a laser speckle field, and by loading a second (heavy) component into the system \cite{exp_weak_1,exp_weak_2,exp_weak_3,exp_weak_4,exp_demarco}.
Bicromatic lattices are produced by superimposing two optical lattices
with the second one being weak and incommensurate with the first one.
This method has been used to study Anderson-type localization
phenomena \cite{exp_weak_2}. Both weakly interacting and strongly
interacting limits have been studied in a speckle potential which is
implemented by scattering a coherent laser light from a rough surface \cite{exp_weak_1,exp_demarco}.

Considerable amount of theoretical effort has been dedicated in the past to
understanding the phases and phase transitions in the system \cite{Freerick, theory_1, theory_2, theory_3, theory_4, theory_5, theory_6, theory_7,
theory_8, theory_9, Weichman}.
In addition to the Mott insulator (MI) and superfluid (SF) states present
in the pure case, in the presence of disorder, the ground state features a third, non-conducting but compressible phase, the Bose glass (BG).
Whether there exist a direct transition between the MI and SF states
has been the subject of a long debate. Fisher {\it et al.} \cite{Fisher} argued that
such a transition is unlikely, i.e. for finite disorder MI and SF phases
are always separated by BG, but alternative possibilities were not
ruled out rigorously. This resulted in a controversy since, on one hand,
mean-field type theories are inadequate in capturing the physics
of rare statistical fluctuations driving the MI-BG transition, and,
on the other hand, various numerical techniques are
severely limited by finite size effects \cite{theory_2, theory_4, theory_5}.
The controversy has been resolved in Refs.~\cite{3d_1,3d_2} which proved the
theorem of inclusions and concluded that all transitions between the fully
gapped and gapless ground states in disordered systems are of the Griffiths
type and thus the resulting gapless phase is insulating. Moreover,
the original conjecture \cite{Fisher,Freerick} that the MI-BG boundary
corresponds to the disorder bound, $\Delta$, equal to the MI gap
in a pure system turns out to be a rigorous result by the same theorem.
Here it is important that disorder is bounded since otherwise the MI phase
is eliminated altogether. [We note that proofs based on rare statistical
fluctuations are valid only in the thermodynamic limit; in finite
experimental systems phase boundaries are replaced by crossovers, including
complete elimination of the BG state for weak enough disorder.]

In the present work we study a two-dimensional disordered Bose-Hubbard model
and present the first accurate results for its ground state phase diagram
at unity filling factor \cite{comment_Lin}. The numerical method of solution is based on the
lattice path integral Monte Carlo using Worm algorithm \cite{Worm}.
We pay special attention to the critical behavior
of the system in the limit of weak disorder in proximity to the SF-MI transition
in the homogeneous system.

The disordered Bose-Hubbard Hamiltonian reads as:
\begin{eqnarray}
\!\!\!\!\!\!
H= -t \sum_{\langle i j \rangle} { a_i^\dag a_j^{\,}} + \frac{U}{2} \sum_{i} {n_i(n_i-1)} + \sum_{i} {(\varepsilon_i\! -\! \mu)n_i}\,  ,
\end{eqnarray}
where $a_i^\dag(a_i^{\,})$ is the boson creation(annihilation) operator,
$\langle i j \rangle$ denotes nearest neighbor sites, and
$n=a_i^\dag a_i$ is the density operator.
In what follows we use the hopping matrix element $t$ as a unit of energy,
$U$ is the on-site repulsion between atoms, and $\mu$ is the chemical potential.
Random disorder, $\varepsilon_i$, is uniformly distributed on the $[-\Delta,\Delta]$ interval and no correlations between the sites of a square lattice.
In our numerical study, we work at filling factor $n=1$ and, for definiteness,
choose the chemical potential in the middle of the MI gap when appropriate,
i.e. gaps for creating particle and hole excitation in the MI phase are both equal to $E_{g/2}$.

To construct the phase diagram we employ standard procedures.
The SF-BG transition lines are determined from finite size scaling of
the superfluid stiffness calculates from the statistics of winding numbers squared,
$\langle W^2 \rangle$, using formula $\Lambda_s=\langle W^2 \rangle L^{2-d}/Td$ \cite{Ceperley}.
According to the theorem of inclusions, the BG-MI line is determined by the
$\Delta=E_{g/2}$ criterion (recall, that it is impossible to detect
this boundary directly in finite-size simulations).

\begin{figure}
\begin{center}
\includegraphics[angle=0,width=\columnwidth]{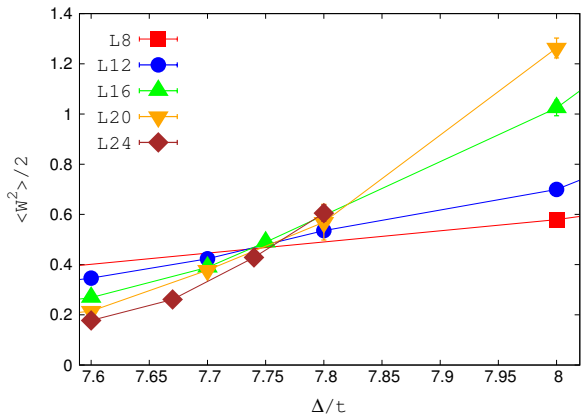}
\end{center}
\vspace*{-1cm}
\caption{(Color online) Finite-size scaling of average winding
numbers squared as a function of disorder bound $\Delta$ for
interaction strength $U/t=22$ and dynamical exponent $z=2$. }
\label{fig:u22}
\end{figure}

Figure~\ref{fig:u22} shows $\langle W^2 \rangle/2$ {\it vs} $\Delta/t$ curves at
$U/t=22$ for different system sizes. In this case we scale space and imaginary time dimensions according to the dynamical critical exponent $z=2$ predicted in Ref.~\cite{Fisher} (strictly speaking, in the thermodynamic limit any value of
$z>0$ can be used for determination of the critical point).
Barely measurable flow of intersection points with the system size
allows us to estimate transition points with relatively high accuracy.
From the data in Fig.~\ref{fig:u22} we deduce $\Delta_c/t=7.76\pm0.04$.
% {\bf We probably should not reproduce our own figure again}
%\begin{figure}
%\begin{center}
%\includegraphics[angle=0,width=\columnwidth]{phase_diagram.pdf}
%\end{center}
%\caption{ The phase diagram $\mu / U$ vs $t/U$ taken from Ref \cite{Fisher}, corresponding to the reentrant behavior of BG in Fig.~\ref{fig:global}  }
%\label{fig:phase_diagram}
%\end{figure}
%%%%%%%%%%%%%%%%%%%%%%%%%%%%%%%%%%%%%%%
%%%%%%%%%%%%%%%%%%%%%%%%%%%%%%%%%%%%%%%%%
%%%%%%%%%%%%%%%%%%%%%%%%%%%%%%%%%%%%%%%
\begin{figure}
\begin{center}
\includegraphics[angle=0,width=\columnwidth]{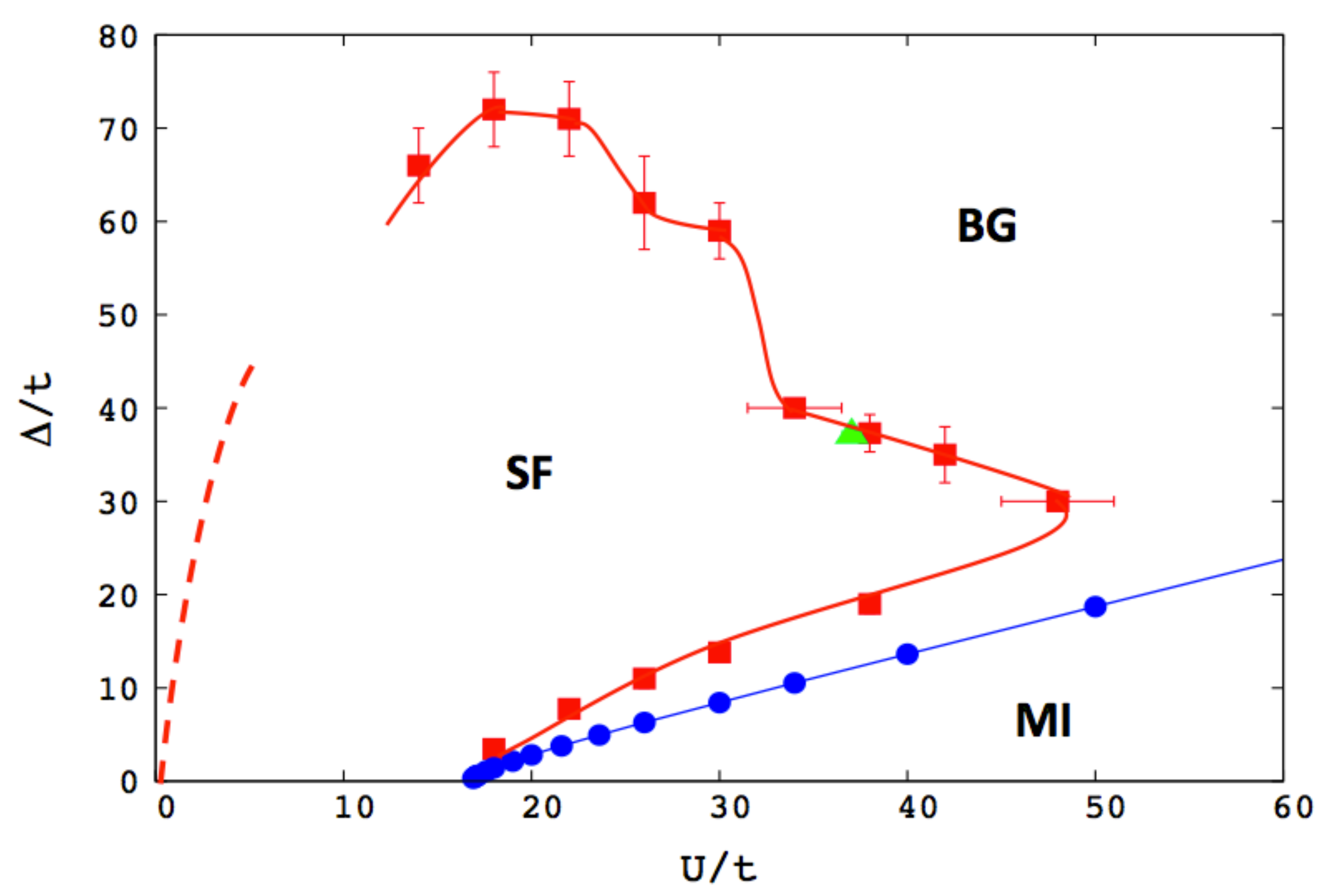}
\end{center}
\vspace*{-0.5cm}
\caption{(Color online). Zero temperature phase diagram of the two-dimensional Bose-Hubbard model at filling factor $n=1$.
 The MI-BG transition at $\Delta=E_{g/2}$ is obtained using gap data from
Ref.~\cite{Capogrosso_Soyler}. (The green triangle is the point on the SF-BG boundary obtained in Ref.~\cite{Lin}.) }
\label{fig:global}
\end{figure}
%%%%%%%%%%%%%%%%%%%%%%%%%%%%%%%%%%%%%%%%%
The full ground state phase diagram in the $(U,\Delta )$ plane
is shown in Fig.~\ref{fig:global}. The boundary between the MI and BG phases
at $\Delta/t=E_{g/2}$ was constructed using data for the MI gap in a pure system calculated in Ref.~\cite{Capogrosso_Soyler}.

Reentrant behavior, similar to the one observed in one- and three-dimensional
phase diagrams, is also present in two dimensions, in agreement with earlier observations \cite{theory_2,Lin}.
In compliance with the theorem of inclusions, BG always separates
SF and MI states which meet only at the MI-SF transition point
of the clean system at $U_c/t=16.7424(5)$.
As we move away in the vertical direction, weak disorder works in favor of the
SF phase shifting the transition points to the right---this appears to be
a common behavior in all physical dimensions. The mechanism, however, is
not universal. In 1D, it can be explained by the destructive interference of the vortex instanton contributions to the partition function \cite{theory_6}.

In the weak interaction limit, $U\to 0$, the transition
line does to zero with an infinite slope,
$\Delta_c \propto \sqrt{U}$ \cite{weak_int_1_2}. In this region, interaction is
very efficient in screening deep potential wells and stabilizing superfluidity.
Numerically, this region is extremely hard to study \cite{lakes}
and we were not able to see its asymptotic laws.

\begin{figure}
\begin{center}
\includegraphics[angle=0,width=\columnwidth]{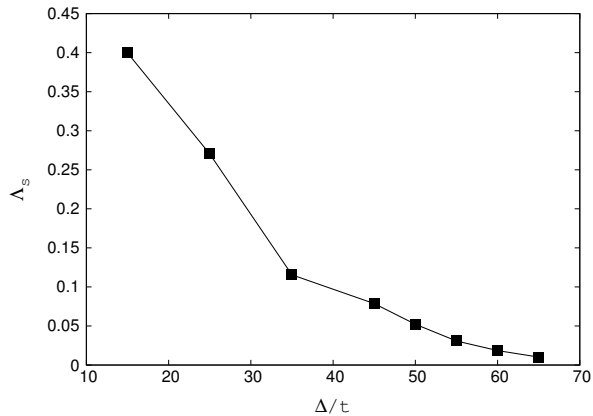}
\end{center}
\vspace*{-1cm}
\caption{Superfluid stiffness as a function of disorder bound $\Delta$ at fixed interaction strength,
$U/t=26$ for system size $L=12\times12$. }
\label{fig:diagonal}
\end{figure}

\begin{figure}
 \begin{center}
\includegraphics[angle=0,width=\columnwidth]{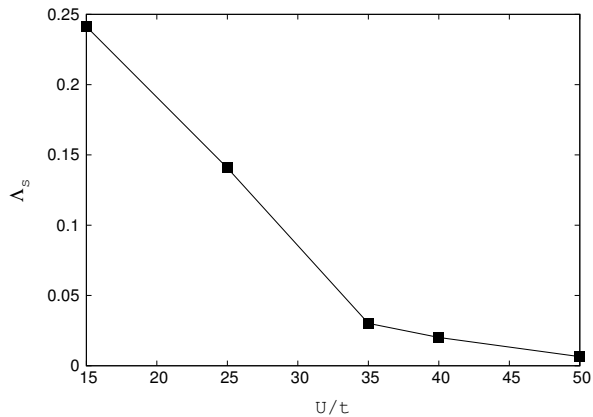}
\end{center}
\vspace*{-1cm}
\caption{Superfluid stiffness as a function of interaction strength $U/t$ at fixed disorder bound,
$\Delta/t=35$ for system size $L=12\times12$. }
\label{fig:parallel}
\end{figure}

Remarkably, superfluidity persists up to extremely large values of
disorder and interactions. For intermediate interactions, superfluidity
survives when disorder potential is about ten times larger
than the bandwidth, $\Delta/t=70$.  Likewise, the superfluid `finger' at
$\Delta/t=30$ extends all the way to $U/t=48$. However, superfluid properties
of the system in the large disorder and large interaction limits are not robust.
In Figs.~\ref{fig:diagonal} and \ref{fig:parallel} we plot the superfluid stiffness
calculated along two representative cuts of the phase diagram, one at fixed
interaction $U/t=26$ and the other at fixed disorder $\Delta/t=35$.
Small values of $\Lambda_s$ for $\Delta > 30t$ ensure small superfluid-normal
transition temperatures according to the Nelson-Kosterlitz relation
$T_c/t = (\pi/2) \Lambda_s(T_c)$. Note that Figs.~\ref{fig:diagonal} and
\ref{fig:parallel} can be used to determine upper bounds on $T_c$
because $\Lambda_s(T_c)<\Lambda_s(0)$.
For example, numerical simulations of the SF-normal transition temperature
at $U/t=26$ and $\Delta/t=35$ yielded $T_c/t=0.07\pm0.012$ below the $T_c/t=0.08$ estimate.

Low transition temperatures have important consequences for current cold-atom
experiments. Though the $(U/t=26, \Delta/t=35)$ point is chosen to be
far from the edges of the SF-BG boundaries in Fig.~\ref{fig:global},
the value of $T_c$ is well below typical experimental
temperatures which are still at (or above) the tunneling amplitude $t$.
Thus observing the ground state phase diagram remains a challenging task.
In current experimental setups, only a small fraction
of the superfluid region will survive finite temperature effects.
%%%%%%%%%%%%%%%%%%%%%%%%%%%%%%%%%%%%%%%
%%%%%%%%%%%%%%%%%%%%%%%%%%%%%%%%%%%%%%%%%
%%%%%%%%%%%%%%%%%%%%%%%%%%%%%%%%%%%%%%%
%%%%%%%%%%%%%%%%%%%%%%%%%%%%%%%%%%%%%%%%%
%%%%%%%%%%%%%%%%%%%%%%%%%%%%%%%%%%%%%%%
\begin{figure}
\begin{center}
\includegraphics[angle=0,width=\columnwidth]{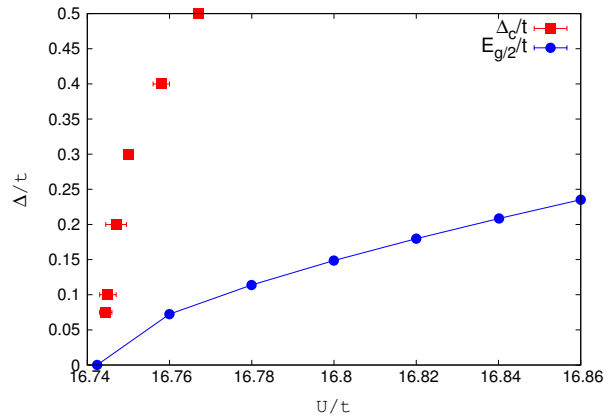}
\end{center}
\vspace*{-1cm}
\caption{(Color online). Phase diagram in the vicinity of the Mott lobe
for weak disorder. MI-BG
transition curve is determined by the MI gap, $\Delta=E_{g/2}$, calculated in
Ref.~\cite{Capogrosso_Soyler}. Note extremely weak
dependence of the SF-BG critical point on disorder for small $\Delta$.}
\label{fig:critical}
\end{figure}
%%%%%%%%%%%%%%%%%%%%%%%%%%%%%%%%%%%%%%%%%

Let us now focus on the weak disorder case at the tip
of the Mott lobe in the pure system. Consider the
Euclidian superfluid hydrodynamic action ($\hbar=1$)
\begin{eqnarray}
S=\int d\vec{r} \int d\tau \left\{ i\langle n \rangle \dot{\Phi} + \frac{\kappa}{2}\dot{\Phi}^2 + \frac{\Lambda_{s}}{2}[\nabla \Phi]^2 \right\}\, ,
\label{action}
\end{eqnarray}
where $\kappa$ is the compressibility, $\tau$ is the imaginary time,
$\Phi (\vec{r},\tau)$ is the superfluid phase field, and
$\langle n (\vec{r}) \rangle $ is the average density at the lattice
point $\vec{r}$ (the integral over $d\vec{r}$ is understood as  a discrete lattice sum,
and the gradient as a finite difference).
Since the first term contains full time derivative
of the phase variable, we observe that
(i) the integral $\int  \dot{\Phi} \, d\tau$ can be only a multiple of $2\pi$,
(ii) in a pure system with $ \langle n (\vec{r}) \rangle =1 $ this term is
irrelevant and the effective action (after rescaling of the time variable
with the sound velocity) is that of the classical 3D XY-model, (iii) in a disordered
system this term is of any importance (not a multiple of $2 \pi$) only
in the presence of topological defects in the phase field. More precisely,
its value is purely imaginary and for the (2+1)-dimensional space-time vortex-loop
with projected (on the space plane) $d$-dimensional algebraic area $A$ is given
by a simple formula
\begin{eqnarray}
 S_{\rm loop} = 2 \pi  i \int_{A} d\vec{r} \, \langle \delta n (\vec{r}) \rangle \, ,
 \label{phase}
\end{eqnarray}
where $\delta n (\vec{r}) = n (\vec{r}) -1$ is the local density fluctuation
about the mean value.

The nature of the SF-MI transition in a clean system described by the 3D XY-model
is linked to the proliferation of large vortex-loop instantons which disorder
the phase field. In the absence/irelevance of the phase term all vortex-loop
instantons act in `unison', in other words their contributions interfere constructively
in the partition function. This is no longer the case in the disordered system since
an area integral in Eq.~(\ref{phase}) is now a random variable. When random vortex
phases become large, their contributions to the partition function cancel each other
making them inefficient in destroying phase coherence across the system.
Let us estimate the typical phase of a vortex-loop of size $\xi$ where $\xi$ is the correlation length. It is proportional to the total particle
number fluctuation in the area $A\sim \xi^2$ in response to the random chemical
potential fluctuation in this region, $\delta \mu _A = A^{-1} \int \epsilon_r d^2r$,
which, by the central limit theorem, scales as $\Delta / \xi $. Using system's compressibility, $\kappa=\partial n/\partial \mu$, we then find
\begin{eqnarray}
{\rm Im}\,  S_{\rm loop} \sim  2\pi  A \kappa \Delta/ \xi \sim 2\pi  \xi \kappa \Delta \;.
\label{phase2}
\end{eqnarray}
As long as disorder remains perturbative (i.e. $S_{\rm loop}\ll 1$), we have
$\xi \sim (1-U_c(\Delta)/U)^{-\nu}$ where $\nu =0.67155$ is the correlation
length exponent and $U_c(\Delta)$ is the critical interaction for a given $\Delta$.
Equation (\ref{phase2}) extrapolated to ${\rm Im}\,  S_{\rm loop} \sim 1$
predicts how close $U$ should be to $U_c(\Delta)$ to start seeing relevance of 
disorder, provided the system size is {\it exponentially} large:  $L >\xi  \propto \exp (U/\Delta) $,
as it follows from the estimate 
\begin{eqnarray}
{\rm Im}\,  S_{\rm loop}\propto  (\Delta /U) \ln \xi 
\label{phase3}
\end{eqnarray}
justified below.

One might think that Eq.~(\ref{phase2}) implies linear scaling of phase
with the correlation length. However, in the vicinity of the particle-hole
symmetric SF-MI transition point $U_c/t$, the renormalized compressibility
itself vanishes as $\kappa \sim 1/U\xi$, canceling dependence on
$\xi$ in Eq.~(\ref{phase2}). It means that each length-scale contributes
equally to the vortex-loop phase, i.e. the final dependence is only
logarithmic, bringing us to Eq.~(\ref{phase3}).

 On the other hand, we conclude that $d=2$ is the `critical dimension' starting from
which the relevance of weak disorder can not be seen by assuming that compressibility
follows the Josephson relation, $\kappa \propto \xi^{1-d}$, meaning that
the shape of the critical line $U_c(\Delta)$ is associated with essentially
non-universal microscopic physics.
[Finite compressibility in Eq.~(\ref{phase2}) clearly implies relevance of disorder
on length scales $ > (\kappa \Delta)^{-2/d}$.]
This is to be contrasted to $d=1$ where the scale of relevance of weak disorder defines the form of the line $U_c(\Delta)$, see \cite{theory_6}.

In Fig.~\ref{fig:critical} we show the phase diagram close to the tip
of the Mott lobe. Critical points were determined using finite-size scaling with
$z=1$. Clearly, critical points for the SF-BG transition are at disorder
values much larger than the $E_{g/2}$ boundary for the MI phase. Surprisingly enough,
the first data points  for the SF-BG line are indistinguishable
from the critical value in the clean system $U_c/t=16.7424(5)$ within the
error bars. This was unfortunate to some degree because even though critical
points were determined with accuracy better than four digits we still
did not have enough parameter range to make an unambiguous case for
the form of the line $U_c(\Delta)$.

Summarizing, we have presented the full ground state phase diagram of the
disordered Bose-Hubbard model at unity filling factor. Interestingly,
while the superfluid phase is remarkably stable against strong interactions
and disorder it is rather fragile with regards to finite temperature effects.
Our  numerical data feature an essentially vertical SF-BG line for weak disorder;
understanding physics behind this phenomenon in $d\ge 2$ remains a challenge.

\vspace*{-5mm}
%%%%%%%%%%%%%%%%%%%%%%%%%%%%%%%%%%%%%%%%%%%%%%%%%%%%%%%%%%%%%%%


\begin{thebibliography}{50}
\vspace*{-5mm}
\bibitem{Giamarchi}
T. Giamarchi and H. J. Schulz, Phys. Rev. \textbf{B 37} 325 (1988).
\bibitem{Fisher}
M. P. A. Fisher \textit{et al.}, Phys. Rev. \textbf{B 40}, 546 (1989).
\bibitem{He4_vycor_1_2_3}
B. C. Crooker  \textit{et al.}, Phys.
Rev. Lett. \textbf{51}, 666 (1983); M. H. W. Chan \textit{et al.}, Phys. Rev. Lett. \textbf{61}, 1950 (1988); J. D. Reppy, J. Low Temp. Phys. \textbf{87}, 205 (1992).
\bibitem{sc_film}
A. M. Goldman and Y. Liu, Physica \textbf{D 83}, 163 (1995).
\bibitem{josephson_junction}
H. S. J. van der Zant \textit{et al.}, Phys. Rev. B \textbf{54}, 10080 (1996).
\bibitem{He_thinfilms_1_2}
P.A. Crowell, F. W. Van Keuls and J. D. Reppy, Phys. Rev. Lett. \textbf{75}, 1106 (1995);
Phys. Rev. \textbf{B 55}, 12620 (1997); G. A. Csathy, J. D. Reppy, and M. H. W. Chan, Phys. Rev. Lett. \textbf{91}, 235301 (2003).
\bibitem{Jaksch}
D. Jaksch \textit{et al.}, Phys. Rev. Lett. \textbf{81}, 3108 (1998).
\bibitem{Greiner}
M. Greiner \textit{et al.}, Nature \textbf{415}, 39 (2002).
\bibitem{exp_weak_1}
J. Billy \textit{et al.}, Nature
\textbf{453}, 891 (2008).
\bibitem{exp_weak_2}
G. Roati \textit{et al.},  Nature \textbf{453}, 895 (2008).
\bibitem{exp_weak_3}
J. E. Lye \textit{et al.},
Phys. Rev. Lett.  \textbf{98}, 130404 (2007).
\bibitem{exp_weak_4}
M. Zaccanti \textit{et al.}, Nature Physics \textbf{6}, 354 (2010).
\bibitem{exp_demarco}
M. White \textit{et al.}, Phys. Rev. Lett.
\textbf{102}, 055301 (2009).
\bibitem{Freerick}
J. K. Freericks and H. Monien, Phys. Rev. \textbf{B 53}, 2691 (1996).
\bibitem{theory_1}
R. T. Scalettar, G. G. Batrouni, and G. T. Zimanyi, Phys. Rev. Lett. \textbf{66}, 3144 (1991)
\bibitem{theory_2}
W. Krauth, N. Trivedi, and D. Ceperley, Phys. Rev. Lett. \textbf{67}, 2307 (1991).
\bibitem{theory_3}
B. Damski \textit{et al.}, Phys. Rev. Lett.
\textbf{91}, 080403 (2003).
\bibitem{theory_4}
R. V. Pai \textit{et al.},
Phys. Rev. Lett. \textbf{76}, 2937 (1996).
\bibitem{theory_5}
M. Makivic, N. Trivedi, and S. Ullah,
Phys. Rev. Lett. \textbf{71}, 2307 (1993).
\bibitem{theory_6}
B. V. Svistunov, Phys. Rev. \textbf{B 54}, 16131 (1996).
\bibitem{theory_7}
U. Bissbort and W. Hofstetter, Europhys. Lett. \textbf{86}, 50007 (2009).
\bibitem{theory_8}
F. Kr\"{u}ger, J. Wu, and P. Phillips,
Phys. Rev. \textbf{B 80}, 094526 (2009).
\bibitem{theory_9}
N. Prokof'ev and B. Svistunov. Phys. Rev. Lett. \textbf{92}, 015703 (2004).
\bibitem{Weichman}
P. B. Weichman, Mod. Phys. Lett. \textbf{B 22}, 2623 (2008).
\bibitem{3d_1}
L. Pollet \textit{et al.}, Phys. Rev. Lett. \textbf{103}, 140402 (2009).
\bibitem{3d_2}
V. Gurarie \textit{et al.}, Phys. Rev.\textbf{B 80} 214519 (2009).
\bibitem{comment_Lin} Very recently, Lin, S{\o}rensen, and  Ceperley presented new numeric data for the two-dimensional disordered Bose-Hubbard model \cite{Lin};
at the unity filling, however, they have only one point (see our Fig.~\ref{fig:global}).
\bibitem{Lin} F.~Lin, E.~S.~S{\o}rensen, and D.~M.~Ceperley, arXiv:1107.3107.
\bibitem{Worm}
N. V. Prokof'ev, B. V. Svistunov, and I. S. Tupiysyn, Phys. Lett. A
\textbf{238}, 253 (1998); Sov. Phys. JETP \textbf{87}, 310 (1998).
\bibitem{Ceperley}
D. M. Ceperley and E. L. Pollock, Phys. Rev. \textbf{B 36},
8343 (1987).
\bibitem{Capogrosso_Soyler}
B. Capogrosso-Sansone \textit{et al.},
Phys. Rev. \textbf{A 81}, 053622 (2010).
\bibitem{weak_int_1_2}
G. M. Falco, T. Nattermann, and V. L. Pokrovsky, Europhys. Lett. \textbf{85}, 30002 (2009);  Phys. Rev. \textbf{B 80} 104515 (2009).
\bibitem{lakes} At $U\to 0$, the system is a set of large percolating superfluid lakes \cite{weak_int_1_2}, the self-averaged macroscopic behavior setting  in only at very large
 scales of distance.



\end{thebibliography}
\end{document}